\documentclass[english,letterpaper,prl,aps,showpacs,superscriptaddress,floatfix,twocolumn]{revtex4-1}
\usepackage[T1]{fontenc}
\usepackage[latin9]{inputenc}
\usepackage{amssymb}
\usepackage{graphicx}
\usepackage{babel}
\usepackage{subeqnarray}
\usepackage{color}
\usepackage{ulem}
\begin{document}

\title{Phonon cooling by an optomechanical heat pump}

\author{Ying Dong}
\affiliation{Department of Physics, Hangzhou Normal University, Hangzhou, Zhejiang 310036, China}
\affiliation{B2 Institute, Department of Physics and College of Optical Sciences, University of Arizona, Tucson, Arizona 85721, USA}
\author{F. Bariani}
\affiliation{B2 Institute, Department of Physics and College of Optical Sciences, University of Arizona, Tucson, Arizona 85721, USA}
\author{P. Meystre}
\affiliation{B2 Institute, Department of Physics and College of Optical Sciences, University of Arizona, Tucson, Arizona 85721, USA}

\begin{abstract}
We propose and analyze theoretically a cavity optomechanical analog of a heat pump that uses a polariton fluid to cool mechanical modes  coupled to a single pre-cooled phonon mode via external modulation of the substrate of the mechanical resonator. This approach permits to cool phonon modes of arbitrary frequencies not limited by the cavity-optical field detuning deep into the quantum regime from room temperature.
\end{abstract}
\maketitle

Quantum optomechanics has witnessed rapid progress in recent years, resulting in the successful cooling of mechanical modes to their ground state of motion with a combination of cryogenic techniques and optical sideband cooling realized in a number of systems~\cite{RMP,Meystre,COBook}. This opens up a broad spectrum of applications ranging from force and field sensing to tests of the foundations of physics, including the exploration of  the elusive boundary between the quantum and the classical world, studies in quantum thermodynamics, and more.

The recent demonstration of parametric coupling of mechanical modes deep in the quantum regime offers the opportunity to explore a number of aspects of multimode phononics, such as the generation of nonclassical states of mechanical motion and the development of phonon interferometry~\cite{MukundPRL,MukundNJP}. While optomechanical sideband cooling \cite{om_cooling1,om_cooling2} is typically applied to a single mechanical mode, in such multimode applications, and more generally in the emerging area of nonlinear phononics~\cite{Abdi2012, Seok2012, Xuereb2012, Agarwal2014, Xuereb2014, Yamaguchi2014} there is much interest in simultaneously cooling two or more modes of relatively arbitrary frequencies~\cite{Metzger2008}.

We show that this can be achieved in a cooling cycle that relies on the properties of the normal modes (polaritons) of optomechanically coupled optical and mechanical fields~\cite{Keye1,Keye2}. Depending on the frequency detuning between the optical and mechanical fields these are redominantly either photonic or phononic. In the photonic limit they are coupled to a thermal reservoir that is essentially at zero temperature, and in the mostly phononic state to a reservoir at the temperature of the mechanical substrate. In that limit they can also be easily mechanically coupled to any other mode of oscillation of the mechanics by forcing an external modulation of the substrate of the mechanical resonator, as was experimentally realized in Ref.~\cite{MukundPRL}. The advantage of this approach is that specific requirements for the optomechanical coupling and frequencies apply only to the two modes (optical and mechanical) comprising the polaritons, with no restrictions on the additional mechanical modes to be cooled.

The proposed cooling cycle uses a precooled polariton mode as a ``polariton fluid'' whose nature is first changed from photon-like to phonon-like by controlling the pump-cavity detuning. When in the phonon-like state it is parametrically coupled to the phonon mode to be cooled for a duration such that an approximate coherent state transfer is achieved between them. The detuning is then adiabatically returned to a value for which the polariton is photon-like.  Its thermalization at the temperature of the photon bath (with $T_a\approx 0$ and thermal occupation $n_a \approx 0$ at optical frequencies~\cite{note}) irreversibly dumps the excitations that it carried away from the mechanics to the environment, thereby completing the extraction of energy from that mode.

This system is reminiscent of a non-equilibrium heat pump \cite{td_book}, with the important difference that instead of changing the temperature of the cooling fluid, we modify its environment via the polariton dispersion relation \cite{polaritoncooling}. The role of the expansion phase of the refrigerant is achieved by changing the polariton fluid from photon-like to phonon-like,  and the compression-like phase by the reverse process . The heat exchange between the mode to be cooled and the fluid is achieved by phonon population transfer, and heat disposal is achieved by cavity dissipation of the polariton fluid in its photon-like form.

\textit{Model.}  We consider a generic arrangement where a single mode of an electromagnetic cavity is coupled optomechanically to a single vibration mode of a mechanical resonator, which in turn can be coupled to a second mechanical oscillation mode via external actuation, with total Hamiltonian
\begin{equation}
H= H_{\rm om}+H_{\rm st} + H_{\rm dis}
\end{equation}
where $H_{\rm om}$ accounts for the radiation pressure coupling between the cavity mode and one of the mechanical modes, $H_{\rm st}$ describes the interaction between the mechanical modes, and $H_{\rm dis}$ accounts for the intracavity field and mechanical oscillators damping of rates $\kappa$ and $\gamma$, with $\kappa \gg \gamma$ typically.

The optomechanical component of the system consists of a cavity mode of frequency $\omega_a$ coupled to a mechanical mode of frequency $\omega_b$. It is driven by an optical field of frequency $\omega_p$ and amplitude $\alpha_{\rm in}$. We assume that the intracavity field is strong enough that it can be described as the sum of a large mean field $\alpha$ and small quantum fluctuations. In a frame rotating at $\omega_p$ the Hamiltonian $H_{\rm om}$ can then be linearized as
\begin{equation}
H_{\rm om}=-\hbar \Delta(t) \hat a^\dagger \hat a +\hbar \omega_b \hat b^\dagger \hat b + \hbar g(\hat b + \hat b^\dagger)(\hat a + \hat a^\dagger)
\label{H1}
\end{equation}
where the bosonic annihilation operators $\hat a$ and $\hat b$ account for the fluctuations of the optical and mechanical mode around their mean amplitudes $\alpha$ and $\beta$. The radiation pressure interaction is quantified by the constant $g = \alpha g_0$, with $g_0$ the single-photon optomechanical coupling strength, and the laser-cavity detuning $\Delta(t)=(\omega_p -\omega_a)(t) -2\beta g_0$ includes the mean radiation-pressure-induced change in resonator length. In steady state and for small damping, we have $\alpha \approx \alpha_{\rm in}/\Delta$ and $\beta \approx -g_0\alpha^2/\omega_b$. We take $\alpha$ to be real, which can be achieved for an appropriate phase of the pump field.

The coupling between the phonon mode `b' at frequency $\omega_b$ and a second vibration mode `c' at frequency $\omega_c$ can be realized by actuating the substrate at a frequency $\omega_s$ close to their frequency difference, $\omega_s \approx \omega_c-\omega_b$~\cite{MukundPRL}. The resulting coupling is described in a rotating frame at the modulation frequency $\omega_s$ and in the rotating wave approximation by a "state transfer" Hamiltonian
\begin{equation}
H_{\rm st}=\hbar \delta \hat{c}^{\dagger}\hat{c}+\hbar \Omega_0(t)(\hat{b}^{\dagger}\hat{c}+\hat{c}^{\dagger}\hat{b})
\label{Hpara}
\end{equation}
where $\hat c$ is the annihilation operator for mode `c', $\delta =\omega_c-\omega_s \simeq\omega_b$, and $\Omega_0$ is the parametric coupling strength of the two modes proportional to the amplitude of oscillations of the substrate.

In the absence of mechanical coupling, $\Omega_0=0$, the system is driven by the familiar linear optomechanical interaction only. We focus on the red-detuned regime $\Delta < 0$, which in general leads to stable dynamics for small decay rates, and perform a Bogoliubov transformation to diagonalize $H_{\rm om}$ in terms of polaritons described by the bosonic annihilation operators $\hat{A}$ and $\hat{B}$ \cite{Keye1,Keye2}. This gives
\begin{equation}
H_{\rm om}=\hbar \omega_A(\Delta) A^{\dagger}A+\hbar \omega_B(\Delta) B^{\dagger}B,
\label{polaritons}
\end{equation}
with frequencies
\begin{equation}
\omega_{A,B}(\Delta) =\left [\frac{\Delta^{2}+\omega_b^2\pm\sqrt{(\Delta^2-\omega_b^2)^{2}-16g^2\Delta\omega_b}}{2}\right ]^{1/2}.
\label{eq:spectrum}
\end{equation}
A plot of  $\omega_{A,B}(\Delta)$ can be found for instance in Ref.~\cite{Keye1}.

The photon-phonon polaritons are coherent superpositions of the cavity field `a' and the mechanical mode `b'~\cite{Aspelmeyer09}. For $\Delta\ll-\omega_b$, the low-energy polariton branch `B', characterized by the bosonic annihilation operator $\hat B$ and the frequency $\omega_B(\Delta)$, describes phonon-like excitations, with $\omega_{B}$ approaching $\omega_b$. In contrast, on the other side of the avoided crossing, $-\omega_b\ll\Delta<0$, and in the weak coupling regime $g/\omega_b \ll 1$, the operator $\hat{B}$ annihilates photon-like excitations of frequency $\omega_B \sim -\Delta$. The opposite holds for the polariton branch `A', which is photon-like for frequencies far red-detuned from $\Delta = -\omega_b$, and phonon-like near cavity resonance.

\textit{Cooling cycle.}  We now explain how the polariton branch `A' can be exploited as a quantum heat pump refrigerant to cool an arbitrary mechanical mode `c' detuned in frequency from the bare mechanical mode `b'.

We assume that the photon and phonon modes `a' and `b' are initially thermalized at temperatures $T_a\approx 0$ and $T_b$~\cite{note2}. The cooling cycle comprises  four steps: (1) adiabatic change in the frequency $\omega_A$ of the polariton fluid, with an effect similar to the expansion step in conventional heat pumps, and loosely called adiabatic `expansion'  in the following for that reason; (2) ``heat exchange'' between the fluid and the mechanical mode to be cooled; (3) adiabatic change in $\omega_A$ to achieve an analog of the compression stage (adiabatic `compression'); and (4) thermalization of the fluid to $T_a \approx 0$.

The `expansion' is realized by adiabatically changing the detuning $\Delta(t)$ from a large negative value $\Delta_i\ll-\omega_b$, where the polariton `A' is photon-like ($\hat{A} \approx \hat{a}$), to a small negative value $-\omega_b \ll \Delta_f<0$ close to 0, where it is phonon-like ($\hat{A} \approx \hat{b}$). The mechanical coupling is absent at this stage, $\Omega_0 = 0$. This transformation conserves the initial thermal populations of the two polariton modes: the polariton `A' remains unpopulated since $T_a \approx 0$ while the polariton `B' stays at the temperature of mode `b'. Since the polaritons have then exchanged their nature, this transformation effectively swaps the thermal population between photon and phonon fluctuations. The duration $\tau_1$ of this step should be slow enough to guarantee the adiabaticity of the evolution, $\tau_1 \gg 1/(2g)$, but fast compared to the photon and phonon damping times, $\tau_1\ll1/\kappa,1/\gamma$.

Once the detuning has reached the value $\Delta_f$ the coupling $\Omega_0(t)$ between the mechanical  modes `b' and `c' is switched on for a duration $\tau_{2}$. (For simplicity we consider a square pulse, $\Omega_0(t) = \Omega_0$.) At this point the polariton `A' is phonon-like and well approximated by the phonon mode `b'. In the absence of other interactions, the Hamiltonian (\ref{Hpara}) accomplishes a perfect quantum state transfer between the modes `b' and `c' for an appropriate interaction time. For modes initially in thermal states the result can be regarded effectively as a "heat exchange" between them. In practice both optomechanical coupling and dissipation are still effective during $\tau_2$, though, so this is only approximately correct. However these additional effects do not change the dynamics significantly for large detunings between the photon and phonon modes and a short enough $\tau_2$, as confirmed by numerical simulations that account for the full dynamics.

We now estimate the mean excitation transferred between the two modes from the state transfer Hamiltonian only, noting that  In this limit, and for $\Omega_0$ constant during $\tau_2$, the dynamics of the two-mode system is governed by the Heisenberg equations of motion
\begin{eqnarray}
\frac{d\hat b}{dt} =  -i(\omega_b \hat b+\Omega_0 \hat c); \; \frac{d\hat c}{dt}  =  -i(\delta \hat c+ \Omega_0 \hat b),
\label{state trans}
\end{eqnarray}
which yield readily
\begin{equation}
\langle \hat N_c\rangle (t)=\langle \hat N_c\rangle (0)+[\langle \hat N_b\rangle (0)-\langle \hat N_c\rangle (0)]\frac{\Omega_0^2}{\Omega^2}\sin^2(\Omega t).
\label{eq:Nc}
\end{equation}
Here $\hat N_j$ is the number operator for mode `j' and $\Omega=[(\omega_b-\delta)^2/4+\Omega_0^2]^{1/2}$ plays the role of an effective Rabi frequency. The maximum exchange of excitation between the two modes occurs after the interaction time $\tau_2=\pi/(2\Omega)$.

The third, compression-like step is an adiabatic change that returns the polariton mode `A' to  its photon-like nature by changing $\Delta_f$ back to $\Delta_i$. As for the previous adiabatic step, this must take place in a time $\tau_3$ slow enough to guarantee adiabaticity, but fast enough that thermal relaxation remains negligible. The mechanical interaction is off again for the remainder of the cycle.

The last step is the thermalization of the now photon-like polariton with its reservoir, effectively at $T_a \approx 0$, over a time $\tau_4 \gg \kappa^{-1}$. Repeating the full cycle allows one in principle to achieve the ground state cooling of mode `c' from room temperature.

\textit{Numerical simulations.}
This intuitive description of the cooling cycle is confirmed by full numerical simulations of the master equation
\begin{equation}
\frac{d\rho}{dt} = -\frac{i}{\hbar}[H_{\rm om} + H_{\rm st},\rho]+\kappa {\cal L}_{\hat a} [\rho]+\gamma \left \{{\cal L}_{\hat b} [\rho]+{\cal L}_{\hat c} [\rho]\right \},
\label{ME}
\end{equation}
that includes all three modes. Dissipation is described via the super-operators
\begin{eqnarray}
{\cal L}_{\hat a} [\rho]&=&(n_a +1)[\hat a^\dagger \hat a \rho - \hat a \rho \hat a^\dagger +{\rm h.c.})\nonumber \\
&&+ n_a[\hat a \hat a^\dagger \rho - \hat a ^\dagger\rho \hat a +{\rm h.c.}),
\end{eqnarray}
and similarly for ${\cal L}_{\hat b} [\rho]$ and ${\cal L}_{\hat c} [\rho]$, where $n_j$ is the average thermal occupation of mode $j$.

\begin{figure}
\begin{centering}
\includegraphics[width=1.0\columnwidth]{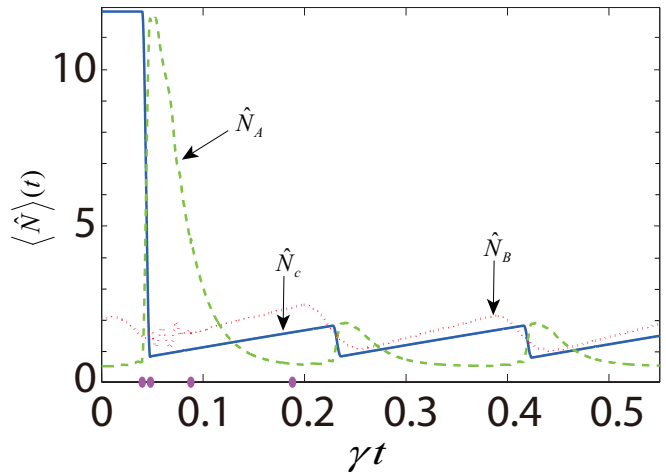}
\caption{Dynamics of the mean polariton populations $\langle \hat N_A\rangle$ (green dashed line) and $\langle \hat N_B\rangle$ (dotted red line) and of the population $\langle \hat N_c\rangle$ of the phonon mode `c' (solid blue line) for a few cooling cycles deep in the quantum regime. Here $\omega_b = \delta = 2 \times 10^{3}$, $\kappa= 40$, $\Delta_i=-6\times 10^3 $, $\Delta_f=-6 \times 10^2$, $g=\Omega_0=2 \times 10^2 $, $\tau_1 = \tau_3 =4 \times 10^{-2}$, $\tau_2=8 \times 10^{-3} $, and $\tau_4= 0.1$, with frequencies in units of the phonon decay rate $\gamma$ and times in units of $\gamma^{-1}$. The initial average populations are $\langle \hat N_A \rangle(0)=0.5$, $\langle \hat N_B\rangle(0)=2$ and $\langle \hat N_c\rangle(0)= 12$. The four dots on the horizzontal axis mark the end of the first 'expansion' step and beginning of `heat exchange', the end of heat exchange and beginning of the `compression' step, the beginning of thermalization, and the end of the first cooling cycle.}
\label{fig1}
\end{centering}
\end{figure}

Results of such a simulation are summarized in Fig.~\ref{fig1}. In this example, the initial phonon number in the mode `c' to be cooled is $\langle \hat N_c\rangle(0) = n_c = 12$, a relatively low value chosen for computational convenience. The polariton mode `A' is photon-like, with $n_a=0$ at optical frequencies., but in this example it is given  the unrealistically large value $\langle \hat N_A\rangle(0) \approx n_a = 0.5$ to illustrate the role of thermal photons. In contrast the polariton mode `B' is phonon-like, with an initial mean phonon number $ \langle \hat N_B\rangle(0) = 2$. It is often the case that the laser-cooled mechanical mode is the lowest frequency one, so we have taken $\omega_c > \omega_b$. The fact that $\langle \hat N_B\rangle(0) < \langle \hat N_c\rangle(0)$ can then be thought of as resulting from the precooling of the phonon mode `b'. Note however that except for its small non-adiabatic and dissipative coupling to mode `A', the polariton mode `B' plays no active role in the operation of the heat pump, so its initial population is of no significant importance.

For short times $t \leq \tau_1$ both $\langle \hat N_A\rangle$ and $\langle  \hat N_c \rangle$ remain essentially constant. During that time $\Delta(t)$ is adiabatically changed from $\Delta_i$ to $\Delta_f$, and the polariton and phonon mode `c' are uncoupled. The adiabaticity of the `expansion' step is confirmed by the fact that $\langle \hat N_A \rangle$ remains approximately constant, with modest non-adiabatic transitions between the two polariton modes.

In the short interval between $\tau_1 < t \leq  \tau_1 + \tau_2$ the interaction between modes `b' and `c' is switched on. Since $\hat{A} \approx \hat{b}$ at $\Delta_f$, we observe an almost perfect "heat transfer" occurring between the the polariton `A' and mode `c'. The mean population of polariton `B' still remains essentially unchanged, which confirms that a simple two-mode model is an excellent approximate description of that step.

The third stroke, of duration $\tau_3$, is a near-adiabatic change of $\Delta$ back to $\Delta_i$. Here non-adiabatic effects are most apparent in oscillations of $\langle \hat N_B\rangle(t)$. This lasts until $\gamma t \approx 0.08$ in Fig.~\ref{fig1}. Finally, heat dissipation into the environment results in the decay of the mean population $\langle \hat  N_A\rangle(t)$ of mode `A', which is now photon-like, to its thermal equilibrium value $n_a$. At the same time, though, the phonon mode `c' is also coupled to a thermal reservoir and consequently $\langle \hat N_c\rangle(t)$ slowly increases. Subsequent cooling cycles permit to keep it at a value near the quantum ground state, as illustrated by the next two cooling cycles in Fig.~\ref{fig1}. (Note that although for the parameters of the simulation $\tau_{1,3} \kappa \approx 1$ optical decay has no detrimental effect during the `expansion' and `compression' stages since it extracts excitations from the system. In addition its contribution to non-adiabatic effects is small~\cite{workmeas}.)

{\it Cooling limit.} To evaluate the cooling limit we first consider the "heat exchange" step. For the small negative detuning $\Delta_f$, the polariton mode `A' consists almost entirely of the phonon mode `b'. Also, the polariton mode `B' is far detuned and remains essentially uncoupled, as we have seen. Under these conditions we can approximate the Hamiltonian (\ref{Hpara}) by
\begin{equation}
H_{\rm st}\approx \hbar \delta c^{\dagger}c+\hbar \Omega'_0(t)(\hat A^{\dagger}c+c^{\dagger}\hat A),
\label{Hpara2}
\end{equation}
a form that results from carrying out the Bogoliubov transformation that diagonalizes $H_{\rm om}$, neglecting the counter-rotating terms in the resulting interaction between the polariton `A' and phonon mode `c', and omitting the coupling to polariton `B' altogether. The effective coupling strength is $\Omega'_0= u \Omega_0$, with $u\approx 1$ for small negative detunings $\Delta_f$ and $g < |\Delta_f|$~\cite{Keye2} is the Bogoliubov transformation coefficient between modes `A' and `b'. We have  confirmed numerically that $\langle  \hat N_c\rangle(t)$ obtained from Eq.~(\ref{Hpara2}) and from a full three-mode analysis coincide almost perfectly for our choice of parameters.

The expression for the mode dynamics obtained from the Hamiltonian (\ref{Hpara2}) has the same analytical form as Eq.~(\ref{eq:Nc}), with $\Omega_0 \rightarrow \Omega'_0$. For the $j^{\rm th}$ cooling cycle and the optimal choice $\Omega' t = \pi/2$, with $\Omega'=[\omega_A (\Delta_f)-\delta]^2/4+\Omega_0^{\prime 2}]^{1/2}$,  the mean population $\langle \hat N_{c, \rm out} \rangle_j$ at the end of the heat exchange step is related to its value $\langle \hat N_{c, \rm in} \rangle_j$ prior to that step by
\begin{equation}
\langle \hat N_{c, \rm out}\rangle_ j= (1-\eta) \langle \hat N_{c, \rm in}\rangle_j  +\eta \langle \hat N_{A, \rm in} \rangle_j,
\label{heat ex}
\end{equation}
where
$\eta=(\Omega_0'/\Omega')^2$
 and $\langle \hat N_{A, \rm in}\rangle_j$ is the mean number of `A' polaritons. If it is properly thermalized while photon-like at the end of the previous cooling cycle we have $\langle \hat N_{A, \rm in} \rangle_j=n_a$, independently of $j$.

We can determine $\langle \hat N_{c, \rm in} \rangle_j$ by noting that between heat exchange steps the phonon mode is decoupled from the polaritons and is only subject to thermalization, so that
\begin{equation}
\langle \hat N_{c, \rm in} \rangle_j=n_c + r (\langle \hat N_{c, \rm out} \rangle_{j-1} - n_c),
\label{j-1}
\end{equation}
where
$r = \exp[-\gamma (\tau_1 + \tau_2+\tau_3 + \tau_4)]$.
Substituting Eq.~(\ref{j-1}) into Eq.~(\ref{heat ex}) and taking the asymptotic limit $\langle \hat N_{c, \rm out}\rangle_{j-1} = \langle \hat N_{c,\rm out}\rangle_j$ gives the cooling limit
\begin{equation}
\langle \hat N_{c,\rm out}\rangle_\infty = \frac{1}{1-r(1-\eta)}\left[ \eta n_a + (1-r)(1-\eta) n_c\right ].
\label{eq:cool_lim}
\end{equation}
The contribution of the phonon thermal noise is fully suppressed for $\eta =1$, which corresponds to the resonance condition $\omega_A(\Delta_f) = \delta$. With $\omega_A(\Delta_f)\approx \omega_b$, this gives $\omega_s = \omega_c - \omega_b$, which is precisely the resonance condition for the substrate modulation frequency to establish the state transfer coupling between the phonon modes `b' and `c'. This ideal case yields the fundamental limit $\langle \hat N_{c, \rm out}\rangle_\infty = n_a$. That is, phonon mode `c' can ideally be cooled to the temperature of the electromagnetic field, $T_a \approx 0$ for visible radiation.

{\it Summary and outlook. }  In conclusion, we have proposed and analyzed a variation of a heat pump that exploits the dispersion relation of a polariton fluid to cool mechanical modes of arbitrary frequency in an optomechanical system.  This can also be understood as a form of reservoir engineering \cite{reseng1,reseng2} where changing the nature of a polariton results in its coupling to reservoirs of different effective temperatures. Importantly, this cooling scheme can readily be extended to a multiplicity of modes by cycling the value of the mechanical coupling with the mode `b' from one to the next.  Figure 2 shows a simulation in which the polariton heat pump is used to cool two mechanical modes `c' and `d'.
\begin{figure}
\begin{centering}
\includegraphics[width=1.0\columnwidth]{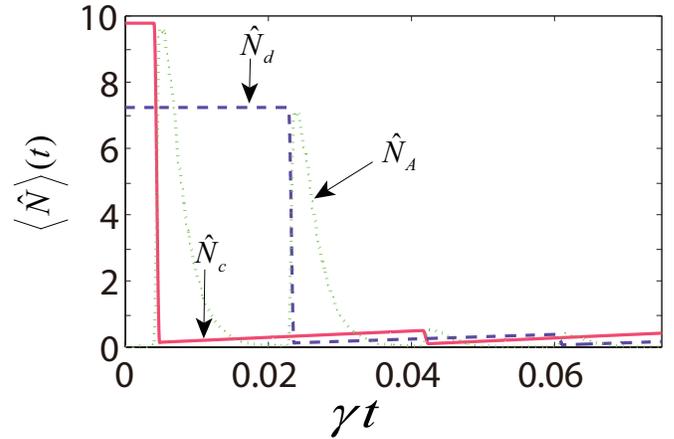}
\caption{Example of two-mode cooling, showing the dynamics of the mean polariton population $\langle \hat N_A\rangle$ (green dotted line) and of the populations $\langle \hat N_c\rangle$ (solid red line) and $\langle \hat N_d\rangle$ (dashed blue line) of the phonon modes `c' and `d'  for two cooling cycles deep in the quantum regime. The parametric coupling between the modes `A' and `c' occurs at times $\gamma t = 0.004$ and 0.0416, and the parametric coupling between 'A' and `d' at times $\gamma t = 0.02228$ and 0.0604. Same parameters as in Fig. 1, except that $\gamma = \kappa/400$. The initial average populations are $\langle \hat N_A \rangle(0)=0$, $\langle \hat N_c\rangle(0)=10$, and $\langle \hat N_d\rangle(0)= 7.4$. The frequencies of modes `c' and `d' are $\omega_c = 1.5\omega_b$ and $\omega_d = 2\omega_b$.}
\label{fig2}
\end{centering}
\end{figure}

Although we have considered numerically only the final stages of cooling, where the mode starts with only a few phonons, the technique  can work in principle starting from room temperature, although the mechanical mode that combines with the optical field to form the polariton fluid does need to be colder. A full quantum simulation of the process starting from room temperature is beyond the capabilities of a numerical simulation, but these first stages of cooling could be described classically straightforwardly, for instance in the framework of a classical Fokker-Planck formalism \cite{fpequation}.

{\it Acknowledgements}. We thank K. Zhang, M. Vengalattore, Y. Patil and S. Singh for useful discussions. This work was supported by the DARPA QuASAR and ORCHID programs through grants from AFOSR and ARO, the U.S. Army Research Office, and the US NSF. YD is supported in part by the NSFC Grants No. 11304072. and the Hangzhou-city Quantum Information and Quantum Optics Innovation Research Team.


\begin{thebibliography}{10}
\bibitem{note2} {The thermal population of mode `b' plays no role in the cycle or in determining its efficiency, as we confirm later, but we nevertheless assume that this mode is sufficiently pre-cooled that its heating rate does not significantly affect the coherent coupling $g$ leading to the polariton picture.}
\end{thebibliography}
\end{document}